\documentclass[12pt, a4paper, twoside]{scrartcl}
\pdfoutput=1

\usepackage[T1]{fontenc}
\usepackage{lmodern}

\usepackage[tracking=true,expansion=true,stretch=15,shrink=15]{microtype}
\DeclareMicrotypeSet*[tracking]{my}
{ font = */*/*/sc/* } 
\SetTracking{encoding = *, shape = sc }{ 45 }
\SetProtrusion{encoding={*},family={bch},series={*},size={6,7}}
{1={ ,750},2={ ,500},3={ ,500},4={ ,500},5={ ,500},
	6={ ,500},7={ ,600},8={ ,500},9={ ,500},0={ ,500}}
\SetExtraKerning[unit=space]
{encoding={*}, family={qhv}, series={b}, size={large,Large}}
{1={-200,-200}, 
	\textendash={400,400}}

\usepackage{setspace}

\usepackage{amssymb}
\usepackage{amsmath}
\usepackage{amsthm}
\usepackage{dsfont}
\usepackage{mathrsfs}
\usepackage{empheq}
\usepackage{bm}
\usepackage{shadethm}
\usepackage{enumerate}
\usepackage{pstricks}
\usepackage{float} 
\usepackage{tikz}
\usepackage[round,sort,comma]{natbib}
\usepackage[automark, headsepline]{scrpage2}
\usepackage{hyperref}
\usepackage{nameref}
\definecolor{TUMblue}{cmyk}{1, .54, .04, .19}
\definecolor{TUMblue2}{cmyk}{0.3, 0.03, 0.07, 0}
\hypersetup{
	colorlinks   = true, 
	urlcolor     = TUMblue, 
	linkcolor    = TUMblue, 
	citecolor   = TUMblue 
}

\usepackage[
left=1.3in,
right=1.3in,
top=0.7in,
bottom=0.7in,
includeheadfoot
]{geometry}

\usepackage[color = TUMblue2]{todonotes}

\newtheoremstyle{new}{12pt}{12pt}{\itshape}{}{\bfseries}{.}{1em}{}
\theoremstyle{new}
\newtheorem{Theorem}{Theorem}
\newtheorem{Corollary}{Corollary}

\newtheorem{Lemma}{Lemma}
\newtheorem{Definition}{Definition}
\newtheorem*{Assumptions}{Assumptions}

\newtheoremstyle{new2}{12pt}{3pt}{}{}{\bfseries}{.}{1em}{}
\theoremstyle{new2}

\newtheorem{Remark}{Remark}

\newcommand{\ind}{\mathds{1}}
\newcommand{\R}{\mathds{R}}
\newcommand{\N}{\mathds{N}}
\newcommand{\Z}{\mathds{Z}}
\newcommand{\E}{\mathrm{E}}

\newcommand{\var}{{\mathrm{Var}}}

\newcommand{\wh}{\widehat}
\newcommand{\wt}{\widetilde}

\allowdisplaybreaks
\begin{document}

	\pagestyle{scrheadings}

	\setcounter{page}{1}
	\pagenumbering{arabic}  
	
	\clearscrheadings
\lohead{T.\ Nagler}
\rohead{\pagemark}
\lehead{Asymptotic analysis of the jittering kernel density estimator}
\rehead{\pagemark}

\title{Asymptotic analysis of the jittering kernel density estimator}

\author{Thomas Nagler\footnote{Corresponding author, Department of Mathematics, Technische Universit{\"a}t M{\"u}nchen, Boltzmanstra{\ss}e 3,  DE-85748 Garching, Germany (email: \href{mailto:thomas.nagler@tum.de}{thomas.nagler@tum.de})}}

\date{\hspace{3pt} \normalsize\today}

\maketitle

\begin{abstract} 
\noindent {\bfseries \sffamily Abstract} \\
\noindent Jittering estimators are nonparametric function estimators for mixed data. They extend arbitrary estimators from the continuous setting by adding random noise to discrete variables. We give an in-depth analysis of the jittering kernel density estimator, which reveals several appealing properties. The estimator is strongly consistent, asymptotically normal, and unbiased for discrete variables. It converges at minimax-optimal rates, which are established as a by-product of our analysis. To understand the effect of adding noise, we further study its asymptotic efficiency and finite sample bias in the univariate discrete case. Simulations show that the estimator is competitive on finite samples. The analysis suggests that similar properties can be expected for other jittering estimators. 
\end{abstract}

	\section{Introduction}

Multivariate density estimation is a central field in nonparametric statistics. Yet many popular methods have a significant drawback in applications: they can only be applied to continuous data. Some estimators have been specifically designed to allow for mixed continuous and discrete data \citep{Ahmad94, Li03, Hall83, Efromovich11}, but the number is small compared to the methods available in a purely continuous framework.

A common trick among practitioners is to make the discrete variables continuous by adding a small amount of noise. The noisy data is continuous and  the usual nonparametric estimators apply. But the addition of random noise can introduce bias, so this procedure generally lacks justification. \citet{Nagler17} showed that adding noise still allows for valid estimates when the noise comes from a certain class of distributions. Then any nonparametric density estimator can be used in the mixed data setting. The resulting estimators are called \emph{jittering estimators}. 

Jittering estimators have so far been neglected in academic research, likely due to the widespread concern that jittering causes a loss in efficiency. The main objective of this article is to demonstrate that this concern is usually unjustified. To this end, we give an in-depth analysis of a simple instance from the class of jittering estimators: the \emph{jittering kernel density estimator}, which is the jittering analog of the classical kernel density estimator \citep{Parzen62,Rosenblatt56,Wand92}. We shall show that it maintains all the properties expected from a good nonparametric density estimator:
\begin{enumerate}[1.]
\item It is asymptotically normal and asymptotically unbiased for discrete variables (\autoref{thm:distribution}).
\item It is strongly and uniformly consistent (\autoref{thm:consistency}).
\item It is relatively efficient, even fully efficient in specific cases (\autoref{sec:efficiency}).
\item It converges at minimax-optimal rates for a large class of target densities (\autoref{thm:upper} and \autoref{thm:lower}). To the best of the author's knowledge, these are the first results on minimax-optimality of nonparametric density estimators for mixed data.
\end{enumerate}
Although focus is on only one instance of the class of jittering estimators, we can expect that others have similar properties.

The remainder of this article is organized as follows. \autoref{sec:estimator} introduces the the jittering estimator and some assumptions. \autoref{sec:asymptotic} gives a comprehensive asymptotic analysis which is complemented by a study of the asymptotic efficiency and finite sample bias in the univariate discrete setting (\autoref{sec:discrete}). \autoref{sec:minimax} establishes with minimax-optimal rates for density estimation in a nonparametric mixed data model. \autoref{sec:simulations} supports demonstrates that the estimator is also competitive on finite samples; \autoref{sec:conclusion} offers conclusions. Proofs of all theorems are deferred to \autoref{appendix}.

    \section{The estimator} \label{sec:estimator}

Suppose that $(\bm Z, \bm X)$ is a random vector with discrete component $\bm Z \in \Z^p$ and continuous component $\bm X \in \R^q$. We explicitly allow for the cases where $p \ge 1$, $q = 0$ (all variables are discrete) and $p = 0$, $q\ge 1$ (all variables are continuous). Our goal is to estimate the density $f$ of $(\bm Z, \bm X)$ based on `observations' $({\bm{ Z}_i, \bm X_i})$, $i = 1, \dots, n$, which are \emph{iid} random vectors having the same distribution as $(\bm Z, \bm X)$. In this context, $f$ is the density with respect to the product of the counting and Lebesgue measures, i.e., \begin{align*}
    f_{\bm Z, \bm X}(\bm z, \bm x) =  \frac{\partial^{q}}{\partial x_{1} \cdots \partial x_{q}} \Pr(\bm{Z} = \bm z, \bm{X} \le \bm x).
\end{align*}

Let $K$ be a real-valued function, called \emph{kernel}, and abbreviate $K(\bm w) = \prod_{j = 1}^k K(w_j)$ for any $\bm w \in \R^k$, $k \in \N$. The classical kernel density estimator is defined as 
\begin{align} \label{eq:kde}
\wh f(\bm z, \bm x) =  \frac{1}{nh_n^p b_n^q} \sum_{i = 1}^n  K\biggl(\frac{\bm Z_{i} - \bm z}{h_{n}}\biggr) K\biggl(\frac{\bm X_{i} - \bm x}{b_{n}}\biggr),
\end{align}
where $h_n, b_n > 0$ are called \emph{bandwidth parameters} and control the amount of smoothing. The above definition of the estimator is simplified to ease our exposition: we use only one parameter ($h_n$) for smoothing all components of $\bm Z$ and one parameter ($b_n$) for smoothing the components of $\bm X$. In practice, one would use a single parameter for each variable or even a bandwidth matrix \citep[see, e.g.,][]{Scott08}. 

The estimator $\wh f$ only works for continuous random vectors. To make it applicable to mixed data, we make all discrete variables continuous by adding noise.
Let $\bm E_i \in \R^p$, $i = 1, \dots, n$, be \emph{iid} random vectors independent from $({\bm{ Z}_i, \bm X_i})$, $i = 1, \dots, n$. Suppose further that the $p$ components of $\bm E_i$ are \emph{iid} with density $\eta$. The jittering kernel density estimator is defined as the classical kernel density estimator applied to $({\bm{ Z}_i + \bm E_i, \bm X_i})$, $i = 1, \dots, n$:
\begin{align} \label{eq:cckde}
\wt f(\bm z, \bm x) =  \frac{1}{nh_n^p b_n^q} \sum_{i = 1}^n  K\biggl(\frac{\bm Z_{i} + \bm E_i -  \bm z}{h_{n}}\biggr) K\biggl(\frac{\bm X_{i} - \bm x}{b_{n}}\biggr).
\end{align}
To facilitate our analysis, the following conditions are imposed on the kernel function:
\begin{Assumptions} \quad \\[-12pt]
\begin{enumerate}
\item[$K1$:] $K\colon [-1, 1] \to \R_{\ge 0}$ is a continuous function satisfying $\int K(t)dt = 1$.
\item[$K2$:] There is $\ell \in \N$, $\ell \ge 2$, such that for $k = 1, \dots, \ell - 1$,
\begin{align*}
\int_{[0, 1]} t^k K(t) dt = 0, \qquad \int_{[0, 1]} t^\ell K(t) dt > 0.
\end{align*}
\end{enumerate}
\end{Assumptions}
\begin{Remark}
A kernel function satisfying K2 is called $\ell$-th order kernel \citep[see, e.g.,][]{Marron94}. \qed
\end{Remark}

We further assume that the noise density $\eta$ belongs to the class $\mathcal{E}_{\gamma_1, \gamma_2}$, as defined in \citet{Nagler17}:
\begin{Definition} \label{def:eta}
We say that $\eta \in \mathcal{E}_{\gamma_1, \gamma_2}$ for some $0 < \gamma_1 \le 0.5 \le \gamma_2 < 1$, if
\begin{enumerate}
\item $\eta$ is an absolutely continuous probability density function,
\item $\eta(x) = 1$ for all $x \in [-\gamma_1, \gamma_1]$,
\item $\eta(x) = 0$  for all $x \in \R \setminus (-\gamma_2, \gamma_2)$.
\end{enumerate}
\end{Definition}
The density of $(\bm Z + \bm E, \bm X)$ is given by
\begin{align*}
f_\eta(\bm z, \bm x) = \sum_{\bm z^\prime \in \Z^p} f(\bm z^\prime, \bm x) \prod_{j = 1}^p\eta(z_j - z^\prime_j), \qquad (\bm z, \bm x) \in \Z^p \times \R^q.
\end{align*}
The class $\mathcal{E}_{\gamma_1, \gamma_2}$ ensures that $f_\eta$ is well-behaved. The most important properties are summarized in the following result \citep[see,][Propositions 2 and 3]{Nagler17}.
\begin{Lemma} \label{lem:eta}
Suppose the components of $\bm E$ are \emph{iid} with density $\eta \in  \mathcal{E}_{\gamma_1, \gamma_2}$. Then the joint density $f_\eta$ of $(\bm Z + \bm E, \bm X)$ satisfies for all $(\bm z, \bm x) \in \Z^p \times \R^q$, and $\bm m \in \N^p$ such that $\sum_{k = 1}^p m_k = \overline m$,
\begin{align*}
	f_\eta(\bm z, \bm x) = f(\bm z, \bm x), \qquad \frac{\partial^{\overline m} f_\eta(\bm z, \bm x)}{\partial z_1^{m_1} \cdots \partial z_p^{m_p}} = 0.
\end{align*}
\end{Lemma}
The first equality implies that we can equivalently estimate $f_\eta$ instead of $f$. This is convenient because $f_\eta$ is the density of a purely continuous random vector. The second equality states that all derivatives w.r.t.\ $\bm z$ vanish, which makes estimation even easier.

\begin{Remark}
The estimator $\wt f$ is similar to the estimators of \citealt{Ahmad94} and \citet{Li03}. The difference lies in the kernel function for discrete data. The estimators of \citealt{Ahmad94} and \citet{Li03} use a deterministic kernel function which is defined on the integers. In contrast, the jittering kernel density estimator \eqref{eq:cckde} uses a random kernel $K\{(\cdot + \bm E_i) / b_n\}$ defined on a compact subset of $\R^p$ where randomness is induced by $\bm E_i$.
\end{Remark}

    \section{Asymptotic analysis in the general setting} 
\label{sec:asymptotic}

\subsection{Asymptotic distribution}

We first study the asymptotic distribution of the jittering kernel density estimator. To motivate our first theorem, we recall a a classical result from kernel density estimation in the purely continuous setting \citep[e.g.,][]{Wand92}. If $f$ is the density of a continuous random vector $(\bm Z, \bm X)$, sufficiently smooth, $\ell = 2$, and $h_n, b_n \to 0$, $nh_n^pb_n^q \to \infty$, then
\begin{align} \label{eq:kde_as}
\begin{aligned}
	\E\bigl\{\wh f(\bm z, \bm x)  \bigr\} &= f(\bm z, \bm x)+ \frac{h_n^2 \sigma_{2}}{2} \sum_{k = 1}^p \frac{\partial^2 f(\bm z, \bm x)}{\partial z_k^2} + \frac{b_n^2 \sigma_2}{2} \sum_{j = 1}^q \frac{\partial^2 f(\bm z, \bm x)}{\partial x_j^2} + o\bigl(h_n^2 + b_n^2\bigr), \\
    \var\bigl\{\wh f(\bm z, \bm x) \bigr \} &= \frac{\kappa^{p + q} f(\bm z, \bm x)}{n   h_n^p b_n^q} + o\biggl(\frac{1}{n h_n^p b_n^q}\biggr), 
\end{aligned}
\end{align}

Recall that $\wt f$ is nothing else than $\wh f$ applied to $({\bm{ Z}_i + \bm E_i, \bm X_i})$, $i = 1, \dots, n$. \autoref{lem:eta} showed that $f_\eta(\bm z, \bm x)$, the density of $({\bm{ Z}_i + \bm E_i, \bm X_i})$, has vanishing derivatives with respect to $\bm z$. We can thus expect the first sum in the bias term in \eqref{eq:kde_as} to vanish asymptotically. In fact, it becomes exactly zero when $h_n \le \min\{\gamma_1, 1 - \gamma_2\}$. The following result improves upon the properties implied by \eqref{eq:kde_as}  by taking these considerations into account. 

\begin{Assumptions} \quad \\[-12pt]
    \begin{enumerate}
        \item[$A1$:] $f(\bm z, \bm x)$ is $\ell + 1$ times continuously differentiable with respect to $\bm x$.
 	    \item[$A2$:] $K1$ and $K2$ hold with $\ell \ge 2$.
        \item[$A3$:] $\eta \in \mathcal{E}_{\gamma_1, \gamma_2}$.
        \item[$A4$:] $b_n \to 0$ and $n h_n^pb_{n}^q \to \infty$ as $n \to \infty$.
        \item[$A5$:] There is $n_0 \in \N$, such that $h_{n} \le \min\{\gamma_1, 1 - \gamma_2\}$ for all $n \ge n_0$.
    \end{enumerate}
\end{Assumptions}

\begin{Theorem} \label{thm:distribution}
	Under assumptions A1-A5, it holds for any  $(\bm z, \bm x) \in \Z^p \times \R^q$,
\begin{align*}
	\E\bigl\{\widetilde f(\bm z, \bm x) \bigr\} &= f(\bm z, \bm x) + \frac{b_n^\ell \sigma_\ell}{\ell!} \sum_{j = 1}^q \frac{\partial^\ell f(\bm z, \bm x)}{\partial x_j^\ell} + o(b_n^\ell), \\
     \var\bigl\{\widetilde f(\bm z, \bm x)\bigr \} &= \frac{f(\bm z, \bm x)}{n b_{n}^q} 
     \bigl\{h_n^{-p} \kappa^{p + q} -  b_{n}^q f(\bm z, \bm x)\bigr\} + o\biggl(\frac{1}{n h_n^p b_{n}^q}\biggr),
\end{align*}
where $\sigma_{\ell} = \int_{-1}^1 s^\ell K(s) ds$ and $\kappa =  \int_{-1}^1  K^2(s) ds$. If further $n h_n^p b_{n}^{q + 2\ell} = O(1)$,
\begin{align*}
\frac{\wt f(\bm z, \bm x) - \E\bigl\{\wt f(\bm z, \bm x)\bigr\}}{\var\bigl\{\wt f(\bm z, \bm x)\bigr\}} \stackrel{d}{\to} \mathcal{N}(0, 1).
\end{align*}
\end{Theorem}

\begin{Remark}
The assumptions in \autoref{thm:distribution} differ from those usually made in the continuous framework. There are no assumptions on the smoothness of $\wh f_{\bm Z + \bm E, \bm X}(\bm z, \bm x)$ with respect to $\bm z$, because its local behavior is controlled by $\eta \in \mathcal{E}_{\gamma_1, \gamma_2}$. Further, $h_n$ is not required to vanish asymptotically, but should be less than $\min\{\gamma_1, 1 - \gamma_2\}$ for large $n$. This is sufficient to ensure that there is no bias with respect to $\bm z$. Further decreasing $h_n$ does not change the bias, but inflates the variance.  \qed
\end{Remark}

\begin{Remark}
The asymptotic variance does not involve on $\eta$ or its class parameters $\gamma_1$ and $\gamma_2$ (and neither does the asymptotic bias). Intuitively, we would expect an increase in the estimator's variance because we are adding random noise. Apparently this effect is dominated by the  sampling variability in the original data and asymptotically negligible. So there should be no benefit from averaging over multiple jitters (at least asymptotically). This is in contrast to empirical processes of jittered data \citep{genest2017asymptotic}.
\end{Remark}


\subsection{Asymptotically optimal bandwidths}
\label{sec:bandwidths}

A standard tool for studying optimal bandwidths is the \emph{asymptotic mean squared error}, 
\begin{align*}
\mathrm{AMSE}\bigl\{\wt f(\bm z, \bm x) \bigr\} = \bigl[\E\bigl\{\widetilde f(\bm z, \bm x) \bigr\} - f(\bm z, \bm x)\bigr]^2 + \var\bigl\{\widetilde f(\bm z, \bm x)\bigr \}.
\end{align*}
Under the assumptions of \autoref{thm:distribution}, we get
\begin{align*}
\mathrm{AMSE}\bigl\{\wt f(\bm z, \bm x) \bigr\} &\approx 
\frac{b_n^{2\ell} \sigma_\ell^2}{(\ell!)^2} \biggl(\sum_{j = 1}^q \frac{\partial^\ell f(\bm z, \bm x)}{\partial x_j^\ell}\biggr)^2 + 
\frac{f(\bm z, \bm x)}{n b_{n}^q}  \bigl\{h_n^{-p} \kappa^{p + q} -  b_{n}^q f(\bm z, \bm x)\bigr\}.
\end{align*}
For $h_n = O(1)$, it is easy to check that the bandwidth $b_n$ minimizing the AMSE satisfies $b_n \sim n^{-1/(2\ell + q)}$. This is well-known as the optimal rate for the classical kernel density estimator when $p = 0$. The AMSE further suggests that it is optimal to choose $h_n$ as large as possible. The largest $h_n$ allowed by A5 is $h_n = \min\{\gamma_1, 1- \gamma_2\}$. Asymptotically, this is the optimal bandwidth. We shall see shortly that this choice means that we are not smoothing the discrete variables at all. This is not unreasonable: in contrast to the continuous case, smoothing discrete variables is not necessary for consistent nonparametric estimation \citep[for a discussion, see,][]{simar2011smooth}.

On finite samples $h_n = \min\{\gamma_1, 1- \gamma_2\}$ can be too small. If $h_n \le  \min\{\gamma_1, 1- \gamma_2\}$, the estimator can be written as 
\begin{align*}
\wt f(\bm z, \bm x) =  \frac{1}{nh_n^p b_n^q} \sum_{i\colon \bm Z_i = \bm z}  K\biggl(\frac{\bm E_i}{h_{n}}\biggr) K\biggl(\frac{\bm X_{i} - \bm x}{b_{n}}\biggr),
\end{align*}
Indeed, the estimator neglects all observations where $\bm Z_i \neq \bm z$ and, thus, does not smooth with respect to the discrete variables. This also means that $\wt f(\bm z, \bm x) = 0$ if $\bm Z_i \neq \bm z$ for all $i  = 1, \dots, n$. \autoref{thm:distribution} implicitly assumes that $n$ is large enough to provide sufficiently many observations with $\bm Z_i = \bm z$. This is guaranteed asymptotically whenever $P(\bm Z = \bm z) > 0$), but often demands sample sizes much larger than what is common. 

We conclude that \autoref{thm:distribution} is not useful for bandwidth selection on samples of small or moderate size. Cross-validation techniques are more appropriate tools in the mixed data setting \citep[see, e.g.,][]{Aitchison76,racine2004nonparametric,hall2004cross}.


\subsection{Consistency} 
\label{sec:consistency}

\autoref{thm:distribution} implies pointwise consistency of the jittering kernel density estimator, but assumption A1 is more strict than necessary. The following result weakens this assumption and additionally establishes strong uniform consistency.

\begin{Assumptions} \quad \\[-12pt]
    \begin{enumerate}
        \item[$A1^\prime$:] The $(\ell-1)$th derivative of $f(\bm z, \bm x)$ exists and is uniformly Lipschitz on  $S \subseteq \Z^p \times \R^q$.
    \end{enumerate}
\end{Assumptions}

\begin{Theorem} \label{thm:consistency}
Suppose that assumptions $A1^\prime$, $A2$--$A5$ hold. Then, for all $(\bm z, \bm x) \in S$,
\begin{align}
\wt f(\bm z, \bm x) - f(\bm z, \bm x) &= O_p\bigl\{b_n^\ell + (nh_n^pb_n^q)^{-1/2}\bigr\}, \label{eq:weak_consistency} \\
\sup_{S} \bigl\vert\wt f(\bm z, \bm x) - f(\bm z, \bm x) \bigr\vert &=  O_{a.s.}\biggl\{b_n^\ell + \biggl(\frac{\max\{\ln \ln n, \ln h_n^{-1}, \ln b_n^{-1}\}}{n h_n^pb_n^q}\biggr)^{1/2}
\biggr\}. \label{eq:strong_consistency}
\end{align}
\end{Theorem}

\begin{Remark}
If there are $h_0 >0, n_0 \in \N$ such that $h_n \in (h_0, \min\{\gamma_1, 1 - \gamma_2\}]$ for all $n \ge n_0$, the rates of convergence in \autoref{thm:consistency} do not involve $p$, the dimension of the discrete variables. So adding more discrete variables does not change the convergence rate of the estimator. In particular, there is no cost for recoding unordered categorical variables into several binary variables.  \qed
\end{Remark}

\begin{Remark}
\begin{enumerate}
\item The best rate in \eqref{eq:weak_consistency} is $n^{-\ell / (2\ell + q)}$ and achieved when $h_n \sim 1$ and $b_n \sim n^{-1 / (2\ell + q)}$. 
\item For $q > 0$, the best rate in \eqref{eq:strong_consistency} is $(n/\ln n)^{-\ell/(2\ell + q)}$ and achieved when $h_n \sim 1$, $b_n \sim (n/\ln n)^{-1/(2\ell + q)}$.  
\item For $q = 0$, the best rate in \eqref{eq:strong_consistency} is $(n/\ln \ln n)^{-1/2}$ and achieved when $h_n \sim 1$. 
\end{enumerate}
\end{Remark}

    \section{A closer look at the univariate discrete setting} 
\label{sec:discrete}

The jittering kernel density estimator $\wt f$ handles continuous variables just like the classical kernel density estimator. How it smooths discrete variables is less obvious. To gain a better understanding, we study its asymptotic efficiency and finite sample bias when there is only one discrete variable ($p = 1$, $q = 0$).

\subsection{Asymptotic efficiency} 
\label{sec:efficiency}

For convenience, set $h_n \equiv \min\{\gamma_1, 1 - \gamma_2\}$. The expectation and variance in \autoref{thm:distribution} become
\begin{align*}
	\E\bigl\{\wt f(z) \bigr\} =  f(z), \quad 
     \var\bigl\{\wt f(z)\bigr \} = \frac{f(z)}{n} 
     \bigl[\min(\gamma_1, 1 - \gamma_2)^{-1} \kappa - f(z)\bigr] + o(n^{-1}),
\end{align*}
The most efficient point estimator for a discrete probability $f(z)$ is the sample frequency $f_n(z) = n^{-1}\sum_{i = 1}^n \ind( Z_i = z)$. It satisfies
\begin{align*}
	\E\bigl\{f_n(z) \bigr\} =  f(z), \quad
    \var\bigl\{f_n(z) \bigr \} = \frac{f(z)}{n}\bigl\{1 - f(z)\bigr\} .
\end{align*}
The \emph{asymptotic relative efficiency (ARE)} of $\wt f$ relative to $f_n$ is defined as 
\begin{align*}
\mathrm{ARE}\bigl\{\wt f(z) : f_n(z) \bigr\} = \frac{\mathrm{AVar}\{f_n(z) \}}{\mathrm{AVar}\{\widetilde f(z) \}},
\end{align*}
where $\mathrm{AVar}$ denotes the leading term of an asymptotic expansion of the variance. The ARE is interpreted as follows: If the estimator $\wt f$ is used with $n$ observations, then one needs $\mathrm{ARE} \times n$ observations to obtain the same accuracy with $f_n$. If the ARE is less than one, then $f_n$ needs less observations, i.e., $f_n$ is more efficient than $\wt f$. If the ARE is greater then one, it is the other way around. If it is exactly one, the two estimators are equally efficient.  

Straightforward calculations yield
\begin{align*}
\mathrm{ARE}\bigl\{\wt f(z) : f_n(z) \bigr\} 
     &= \frac{1 - f(z)}{\min\{\gamma_1, 1 - \gamma_2\}^{-1} \kappa - f(z)}  \\
    & =  \biggl(1 + \frac{\min\{\gamma_1, 1 - \gamma_2\}^{-1} \kappa - 1}{1 - f(z)}\biggr)^{-1}
     \le 1.
\end{align*}
The relative efficiency depends on three quantities:
\begin{itemize}
\item It is increasing in $\min\{\gamma_1, 1 - \gamma_2\}$ and the most efficient choice is $\gamma_1 = \gamma_2 = 1/2$, which corresponds to the uniform error density on $(-1/2, 1/2)$. On the other hand, the relative efficiency approaches 0 for $\gamma_1 \to 0$ or $\gamma_2 \to 1$.

\item It is decreasing in $\kappa$, which is the roughness of the kernel $K$. The `least rough` kernel is the is the uniform kernel, i.e., $K(x) = 2^{-1}\ind(\vert x \vert \le 1)$, for which $\kappa = 1/2$. But this kernel is rather unpopular in practice. A more widely used kernel is the Epanechnikov kernel, $K(x) = 3/4 (1 - x^2) \ind(\vert x \vert \le 1)$, for which $\kappa = 0.6$.

\item It is decreasing in $f(z)$. The worst case is that $f(z) = 1$, for which the ARE is  zero. For a $\mathrm{Bernoulli}(1/2)$ variable, $\mathrm{Uniform}(-1/2, 1/2)$ noise, and the Epanechnikov kernel, we get $\mathrm{ARE} \approx 0.71$.
\end{itemize}

\begin{Remark} \label{kde:eff_ex}

Suppose $\eta$ is the uniform density on $(-1/2,1/2)$ (for which $\gamma_1 = \gamma_2 =1/2$), $h_n = 1/2$, and $K$ is the uniform kernel (for which $\kappa = 1/2$).  Then, the two estimators are equally efficient. In fact, the estimator $\wt f$ becomes
\begin{align*}
	\wt f (z) &= \frac 1 {n h_n} \sum_{i = 1}^n  2^{-1} \ind(\vert Z_{i} + E_{i} - z \vert \le h_n) \\
 &= \frac {2}{n } \sum_{i = 1}^n 2^{-1} \ind(\vert Z_{i} + E_{i} - z \vert \le 1/2) \\ 
 &= \frac 1 n \sum_{i = 1}^n  \ind( Z_i = z),
\end{align*}
which is exactly the sample frequency estimator $f_n$. \qed
\end{Remark}


\subsection{Finite sample bias}
\label{sec:bias}

Assuming  $h_n \le \min\{\gamma_1, 1 - \gamma_2\}$, \autoref{thm:distribution} shows that $\wt f$ is unbiased in a purely discrete setting. On small samples, it is often necessary to choose a larger bandwidth (see \autoref{sec:bandwidths}). When $h_n > \min\{\gamma_1, 1 - \gamma_2\}$, the estimator $\wt f$ is usually biased. 

\begin{Lemma} \label{prop:bias}
Suppose that $\eta \in \mathcal{E}_{\gamma_1, \gamma_2}$ and $K$ satisfies $K1$--$K2$. Then,
\begin{align*}
&\E\bigl\{\widetilde f(z) \bigr\} - f(z) \\
=& \sum_{k = 1}^{\lceil h_n - 1/ 2 \rceil} \frac{\rho_k^\eta(h_n)f(z + k) - \bigl\{\rho_k^\eta(h_n) + \rho_{-k}^\eta(h_n)\bigr\} f(z) + \rho^\eta_{-k}(h_n)f(z - k)}{k^2},
\end{align*}
where $\rho_k^\eta(h_n) = k^2 \int_{A_k^\eta(h_n)} K(t)\eta(k - h_nt) dx$ and 
\begin{align*}
A_k^\eta(h_n) = \bigl[(1 - \gamma_2 - k)h_n^{-1}, \, (-1 + \gamma_2 - k)h_n^{-1}\bigr] \cap [-1, 1].
\end{align*}
\end{Lemma}
To interpret the bias, it is helpful to focus on a simple case first.
 
\begin{Corollary} \label{cor:bias}
Suppose that $\eta(x) = \ind(\vert x \vert \le 1/2)$ and $K$ is a symmetric function satisfying $K1$--$K2$. Then for all $z \in \Z$,
\begin{align*}
\E\bigl\{\widetilde f(z) \bigr\} = f(z) + \sum_{k = 1}^{\lceil h_n - 1/ 2 \rceil} \rho_k(h_n) \Delta^2_k f(z),
\end{align*}
where 
\begin{align*}
\Delta^2_k f(z) =\frac{f(z + k) - 2 f(z) + f(z - k)}{k^2},
\end{align*}
and $\rho_k(h_n) = k^2\int_{A_k(h_n)} K(t) dt$ with
\begin{align*}
A_k(h_n) = \bigl[(1/2 - k)h_n^{-1}, \, (-1/2 - k)h_n^{-1}\bigr] \cap [-1, 1].
\end{align*}
\end{Corollary}

The operator $\Delta^2_k$ is known as the \emph{second order central difference operator} \citep[e.g.,][]{monahan2011numerical}. It is commonly used as numerical approximation of second order derivative of real-valued functions, which is
\begin{align*}
\frac{d^2 f(x)}{d x^2}  = \lim_{s \to 0} \frac{f(x + s) - 2 f(x) + f(x - s)}{s^{2}}.
\end{align*}
We can interpret $\Delta^2_k f$ as a discrete analogue to the second order derivative of a real-valued function. In this aspect, the discrete setting is similar to the continuous one (where the bias of $\wt f$ is proportional to the second order derivative). 

The parameter $k$ is called the \emph{step size} and determines how local the derivative approximation is. The bias of $\wt f$ is a weighted sum of such `derivatives' for several values of $k$. The bandwidth $h_n$ limits the maximal step size and thereby controls the locality of the bias. Although not universally true, smaller values of $h_n$ typically correspond to a smaller bias. A simple counter example is when $f(z + k) = f(z - k) = f(z)$ for all $k \le \lceil h_n - 1/2 \rceil$, where the bias is zero for all $h_n^\prime \le h_n$. There are also situations where decreasing $h_n$ leads to a larger bias. This phenomenon also exists in the continuous setting, but is disguised by asymptotic approximations. When $h_n \le 1/2$ as in \autoref{thm:distribution}, the estimator is unbiased.

The bias in \autoref{prop:bias} can be interpreted similarly. But $\Delta_k^2$ is replaced by a weighted approximation of the derivative. If $\eta$ or $K$ are asymmetric, different weights will be assigned to the `forward derivative' $k^{-1}\{f(z + k) - f(z)\}$ and the `backward derivative' $k^{-1}\{f(z - k) - f(z)\}$.


\section{Minimax rate optimality} \label{sec:minimax}

The \emph{maximum risk} associated with a class of densities $\mathcal{F}$ and a (semi-) distance $d$ is defined as 
\begin{align} \label{problem:maximum_risk_eq}
\mathcal{R}_n(\wh f, \mathcal{F}, d) =  \sup_{f \in \mathcal{F}} \E_f\bigl\{d^2(\wh f, f)\bigr\},
\end{align}
We consider two semi-distances that relate to pointwise and uniform consistency of $\wh f$, respectively:
\begin{align*}
 d_{(\bm z, \bm x)}(\wh f, f) =& \bigl\vert \wh f(\bm z, \bm x) - f(\bm z, \bm x) \bigr\vert, \quad \mbox{for some } (\bm z, \bm x) \in \Z^p \times \R^q, \\
 d_{\infty, \mathcal{S}}(\wh f, f) =& \sup_{\mathcal{S}} \bigl\vert \wh f(\bm z, \bm x) - f(\bm z, \bm x) \bigr\vert, \quad \mbox{for some } \mathcal{S} \subset \Z^p \times \R^q.
\end{align*}
For $\mathcal{F}$, we shall consider all bounded density functions whose continuous part belongs to a H\"older class. For $\bm a \in \N_0^q$, we use the multi-index notations $\vert\bm a \vert = \sum_{j = 1}^q a_j$, $\bm x^{\bm a} = x_1^{a_1}  \cdots  x_q^{a_q},$ and denote the partial derivatives of $f$ with respect to $\bm x$ as
\begin{align} \label{eq:Da}
D_{\bm x}^{\bm a} f(\bm z, \bm x) = \frac{\partial^{\vert \bm a \vert} f(\bm z, \bm  x)}{\partial^{a_1} x_1 \cdots \partial^{a_q} x_q}.
\end{align}

\begin{Definition} \label{problem:H_def}
For $\lambda < \infty$ and $\beta = r + \alpha$,  $r \in \N_0$, $0 < \alpha \le 1$, the class $\mathcal{H}(\beta, \lambda)$ is defined as all functions $f\colon \Z^p \times \R^q \to \R$ such that for all $\bm a \in \N_0$ with $\vert \bm a \vert \le r$,
\begin{enumerate}
\item $f$ is a probability density on $ \Z^p \times \R^q$,
\item $D_{\bm x}^{\bm a} f(\bm z, \bm x)$ exists for all $(\bm z, \bm x) \in  \Z^p \times \R^q $ and
\begin{align*}
\displaystyle \sup_{\bm z \in \Z^p, \bm x, \bm x^\prime \in \R^q} \biggl\{ \frac{\bigl\vert D_{\bm x}^{\bm a} f(\bm z, \bm x) - D_{\bm x}^{\bm a} f(\bm z, \bm x^\prime) \bigr\vert}{\|\bm x- \bm x^\prime \|_2^{\alpha}} + f(\bm z, \bm x) \biggr\} \le \lambda
\end{align*}
\end{enumerate}
\end{Definition}

\begin{Remark}
If $p\ge 1$ and $q = 0$, $\mathcal{H}(\beta, \lambda)$ contains all densities on $\Z^p$. If $p = 0$ and $q \ge 1$, it is a H\"older class on $\R^q$. \qed
\end{Remark}

The following result establishes convergence rates of the jittering kernel density estimator with respect to the maximum risk.

\begin{Theorem} \label{thm:upper}
Denote $\wt f$ as the estimator defined in \autoref{eq:cckde}. Suppose $f \in \mathcal{H}(\beta, \lambda)$ and assumptions A2--A4 of \autoref{thm:distribution} hold with $\ell \ge r + 1$, $\beta = r + \alpha$, $0 < \alpha \le 1$,  $\lambda < \infty$. Assume further that there are $h_0 >0, n_0 \in \N$ such that $h_n \in (h_0, \min\{\gamma_1, 1 - \gamma_2\}]$ for all $n \ge n_0$. Then there exists $\overline c > 0$ such that
\begin{align*}
\limsup_{n \to \infty} r_n^{-2} \mathcal{R}_n(\wh f, \mathcal{F}, d) \le \overline  c, 
\end{align*}
in each of the following cases:
\begin{enumerate}
\item $r_n = n^{-\beta/(2\beta + q)}$, $d = d_{(\bm z, \bm x)}$ 
\item $r_n = (n/\ln n)^{-\beta/(2\beta + q)}$, $d = d_{\infty, \mathcal{S}}$, $q \ge 1$,
\item $r_n = n^{-1/2}$, $d = d_{\infty, \mathcal{S}}$, $q = 0$, $\vert \mathcal{S} \vert < \infty$,
\item $r_n = (n/\ln \ln n)^{-1/2}$, $d = d_{\infty, \mathcal{S}}$, $q = 0$, $\vert \mathcal{S} \vert = \infty$,
\end{enumerate}
for arbitrary $(\bm x, \bm z)  \in \Z^p \times \R^q$ and  $\mathcal{S} \subset \Z^p \times \R^q$. 
\end{Theorem}

We shall see that the rates  in  \autoref{thm:upper} (i)--(iii) are optimal in a minimax sense. The minimax risk is defined as 
\begin{align*}
\mathcal{R}_n^*(\mathcal{F}, d) =   \inf_{\wh f} \mathcal{R}_n(\wh f, \mathcal{F}, d) = \inf_{\wh f} \sup_{f \in \mathcal{F}} \E_f\bigl\{d^2(\wh f, f)\bigr\},
\end{align*}
where the infimum is taken over all possible estimators $\wh f$ of $f$. In our context, an `estimator' is any measurable function of $({\bm{ Z}_i, \bm X_i})$, $i = 1, \dots, n$.

\begin{Definition}
A sequence of positive real numbers $r_n$ is called  
\begin{enumerate}
  \item an upper bound on the minimax rate if there is $\overline c$ such that
  \begin{align*} 
    \limsup_{n \to \infty}  r_n^{-2} \mathcal{R}_n^*(\mathcal{F}, d) \le \overline c. 
  \end{align*}

  \item a lower bound on the minimax rate if there is $\underline c > 0$ such that
  \begin{align*} 
    \liminf_{n \to \infty}  r_n^{-2} \mathcal{R}_n^*(\mathcal{F}, d) \ge \underline c, 
  \end{align*}

  \item a minimax-optimal rate of convergence if both (i) and (ii) hold.
  \end{enumerate}
\end{Definition}

In a purely continuous setting, optimal rates have long been established \citep{Stone80,Stone83,ibragimov1983}. To the best of the author's knowledge, there are no results on optimal rates in the mixed data setting.

To show that a rate is minimax-optimal, we have to check that it is both an upper and lower bound on the minimax rate. \autoref{thm:upper} already gives us an upper bound, since, for any estimator $\wh f$,
\begin{align*}
\mathcal{R}_n^*(\mathcal{F}, d) = \inf_{\wh f} \mathcal{R}_n(\wh f, \mathcal{F}, d) \le \mathcal{R}_n(\wh f, \mathcal{F}, d).
\end{align*}
Lower bounds on the minimax rate can be deduced easily by considering subsets of $\mathcal{H}(\beta, \lambda)$ for which lower bounds are known (see \autoref{appendix:th4}). 

\begin{Theorem} \label{thm:lower}
Let $\mathcal{S} \subset \Z^p \times \R^q$  and $(\bm z, \bm x) \in \mathcal{S}$. The minimax-optimal rate of convergence $r_n^*$ associated with the class $\mathcal{H}(\beta, \lambda)$ and distance $d$ satisfies
\begin{enumerate}[(i)]
\item $r_n^* = n^{-\beta/(2\beta + q)}$, for $d = d_{(\bm z, \bm x)}$ 
\item $r_n^* = (n/\ln n)^{-\beta/(2\beta + q)}$, for $d = d_{\infty, \mathcal{S}}$, $q \ge 1$,
\item $r_n^* = n^{-1/2}$, for $d = d_{\infty, \mathcal{S}}$, $q = 0$, $\vert \mathcal{S} \vert < \infty$,
\item $r_n^* \in [n^{-1/2}, (n/\ln \ln n)^{-1/2}]$, for $d = d_{\infty, \mathcal{S}}$, $q = 0$, $\vert \mathcal{S} \vert = \infty$,
\end{enumerate}
\end{Theorem}

\begin{Remark}
\autoref{thm:upper} and \autoref{thm:lower} imply that the jittering kernel density estimator converges at minimax-optimal rates for cases (i)--(iii). \qed
\end{Remark}

\begin{Remark}
\autoref{thm:lower} only provides an interval for the optimal rate in case (iv). Minimax analysis for this setting is surprisingly har; see \citep{han2015} for minimax rates with respect to the $\ell_1$ distance. The interval is quite narrow, differing only by a factor of size $\ln \ln n$. The exact rate, however, remains an open problem. \qed
\end{Remark}

    \section{Simulation experiments} 
\label{sec:simulations}

The jittering kernel density estimator has appealing asymptotic properties. This may come as a surprise: since we are adding noise to the data, we could expect that the data become less informative and uncertainty increases. We complement our asymptotic arguments with a small numerical experiment that illustrates the small sample performance of the estimator. Because of its wide use and close resemblance to our approach, we will use the estimator of \citet{Li03} as a benchmark.

We use the following setup:
\begin{itemize}
\item We compare three estimators 
\begin{enumerate}
\item \texttt{jkde}: the jittering kernel density estimator with noise density $\eta(x) = \ind(\vert x \vert < 1/2)$, for which $\gamma_1 = \gamma_2 = 1/2$.
\item \texttt{jkde2}: the jittering kernel density estimator with noise density $\eta(x) = f_{U_{1/4, 5}}(x)$ \citep[as in,][Example 3]{Nagler17}, for which $\gamma_1 = 3 / 8$, $\gamma_2 = 5/8$.
\item \texttt{liracine}: the estimator of \citet{Li03}  as implemented in the \texttt{np} package \citep{np}.
\end{enumerate}
Contrary to \eqref{eq:cckde}, we use one bandwidth parameter for each variable. Both estimators use likelihood cross-validation for bandwidth selection. 

\item We estimate the density $f$ of a vector $(\bm Z, \bm X) \in \Z^p \times \R^q$, where $Z_j \sim \mathrm{Binomial}(m, 0.3)$ for all $j =1, \dots, p$, $X_j \sim \mathcal{N}(0, 1)$ for all $j = 1, \dots, q$. For sake of simplicity, all variables are simulated independently.

\item Results are based on $N_{\mathrm{sim}} = 1000$ simulated data sets with sample sizes $n = 50, 200$. 

\item As a performance measure we use the \emph{root average square error (RASE)} computed over a grid in $\Z^p \times \R^q$. More specifically, we use $\mathcal{Z} = \{0, \ldots, m\}$, $\mathcal{X} = \{-2, -1.6, \ldots, 2\}$, and
\begin{align*}
\mathrm{RASE}\bigl(\wh f, f\bigr) = \sqrt{\sum_{z_1 \in \mathcal{Z}} \cdots \sum_{z_p\in \mathcal{Z}} \sum_{x_1 \in \mathcal{X}} \cdots \sum_{x_q \in \mathcal{X}} \bigl\{\wh f(\bm z, \bm x) - f(\bm z, \bm x)\bigr\}^2}.
\end{align*}
\end{itemize}

\begin{figure}[t]
\includegraphics[width=\textwidth]{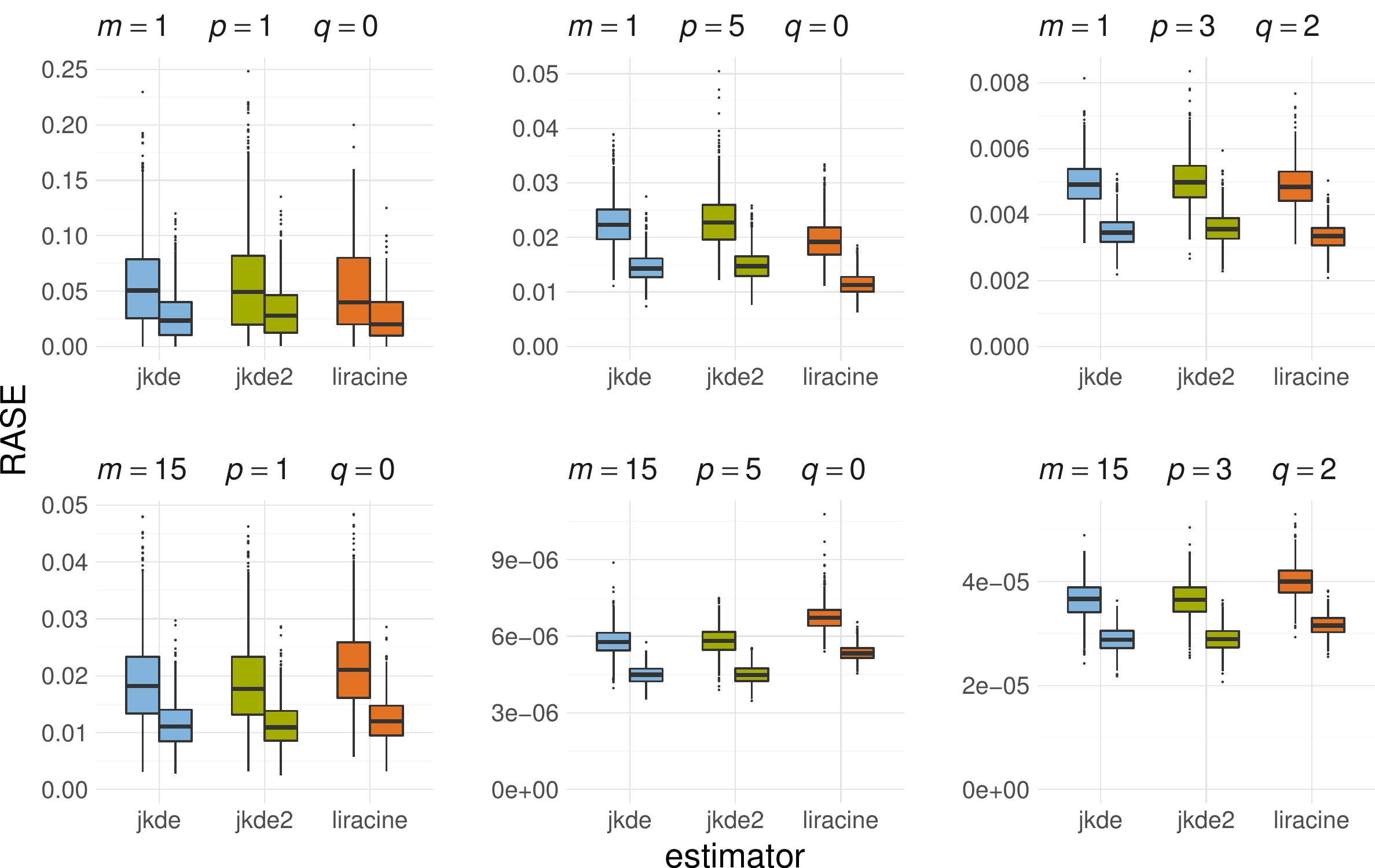}
\caption{RASE achieved by the two estimators for various choices of $p$, $q$, and $m$.  Each estimator is represented by two boxes; the left box corresponds to $n = 50$, the right to $n = 200$.}
\label{sim:results_fig}
\end{figure}

\autoref{sim:results_fig} shows the estimators' performance for various values of $p$, $q$ and $m$. Each estimator is represented by two boxes, where the left box corresponds to $n=50$ and the right box to $n = 200$. The choice of noise density seems to be of minor importance: \texttt{jkde} and \texttt{jkde2} give almost identical results. Compared to \texttt{liracine}, the two estimator show only subtle differences. The two jittering estimators are more accurate in all scenarios with $m = 15$, and less accurate when $m = 1$. This is related to our observation from \autoref{sec:efficiency} that the efficiency is worse when $f(z)$ is large.  The relative performance of the three estimators is consistent across the two sample sizes under consideration. Overall, the jittering estimators are competitive with the benchmark estimator \texttt{liracine}. We found no evidence that adding artificial noise negatively affects the accuracy of the estimates. This confirms what was suggested by the estimator's asymptotic properties.

    \section{Conclusion} \label{sec:conclusion}

This article gave an in-depth analysis of the behavior of the jittering kernel density estimator. It was shown to have appealing large-sample properties and perform well on small samples. 

Although our focus was on a particular instance of the class of jittering estimators, we also learned something about the class as a whole. Adding noise to discrete variables does not have a negative impact on estimation accuracy. This is true for both large samples (as confirmed by our asymptotic analysis) and small samples (as illustrated by simulations). More specifically, it allows for estimators that are optimal in terms of convergence rates and efficiency. It is likely that these findings generalize to more sophisticated density estimators or estimators of functionals of the density, such as regression functions.

\subsection*{Supplementary material}

\begin{itemize}
\item \href{https://github.com/tnagler/cctools}{https://github.com/tnagler/cctools}: an R package implementing the jittering kernel density estimator and likelihood cross-validation for the bandwidths.
\item \href{https://gist.github.com/tnagler/786465cee2c774a844ff1846e7cdacd8}{https://gist.github.com/tnagler/786465cee2c774a844ff1846e7cdacd8}: code for the simulation study in \autoref{sec:simulations}.
\end{itemize}

\subsection*{Acknowledgements}

This work was partially supported by the German Research Foundation (DFG grant CZ 86/5-1).
	
    \appendix

\section{Proofs} \label{appendix}

\subsection{Proof of Theorem 1}  \label{appendix:th1}

We first calculate the bias term. Using a change of variables, we get
\begin{align*}
\E\bigl\{ \wt f(\bm z, \bm x) \bigr\} &=  \frac{1}{h_n^pb_n^q}\E\biggl\{   K\biggl(\frac{\bm Z + \bm E - \bm z}{h_{n}}\biggr)  K\biggl(\frac{\bm X -  \bm x}{b_{n}}\biggr)  \biggr\}\\
&= \frac{1}{h_n^p b_n^q} \int_{\R^{p + q}}   K\biggl(\frac{\bm s - \bm z}{h_{n}}\biggr)  K\biggl(\frac{\bm t -  \bm x}{b_{n}}\biggr) f_\eta(\bm s, \bm t) d \bm s d\bm t \\
&= \int_{\R^{p + q}} K(\bm u)  K(\bm v) f_\eta(\bm z + h_n \bm u , \bm x + b_n \bm v ) d \bm u d\bm v  
\end{align*}
Since $\eta \in \mathcal{E}_{\gamma_1, \gamma_2}$, it holds for all $(\bm z, \bm x) \in \Z^q \times \R^q$ and $0 \le \epsilon \le \min\{\gamma_1, 1 - \gamma_2\}$ that $f_\eta(\bm z + \epsilon , \bm x)  = f(\bm z , \bm x)$. Furthermore, $K$ is zero outside of $[-1, 1]$. Hence, for  $h_n \le \min\{\gamma_1, 1 - \gamma_2\}$,
\begin{align}
\E\bigl\{ \wt f(\bm z, \bm x) \bigr\} &= \int_{\R^{p + q}} K(\bm u)  K(\bm v) f(\bm z, \bm x + b_n \bm v ) d \bm u d\bm v  \notag \\
&= \int_{[-1, 1]^q} K(\bm v) f(\bm z, \bm x + b_n \bm v ) d\bm v. \label{eq:kde_bias}
\end{align}
Recall the derivative notation from \eqref{eq:Da}. An $\ell$-th order Taylor expansion of $f$ yields that
\begin{align*}
\E\bigl\{ \wt f(\bm z, \bm x) \bigr\} - f(\bm z, \bm x) 
&= \sum_{1 \le \vert \bm a \vert  \le \ell} \frac{b_n^{\vert \bm a \vert}}{|\bm a|!} \int_{[-1, 1]^q} K(\bm v) \bm v^{\bm a} D^{\vert \bm a \vert}_{\bm x} f(\bm z, \bm x) d\bm v \\
&\phantom{=} + \sum_{\vert \bm a \vert  = \ell + 1} \frac{b_n^{\ell + 1}}{(\ell + 1)!} \int_{[-1, 1]^q} K(\bm v) \bm v^{\bm a} D^{\vert \bm a \vert}_{\bm x} f(\bm z, \bm x + \tau_{\bm a} \bm v) d\bm v 
\\
= &\phantom{=}  \frac{b_n^{\ell}}{\ell !}\sum_{j = 1}^q \int_{[-1, 1]} K(v_j) v_j^{\ell} \frac{\partial^{\ell} f(\bm z, \bm x)}{\partial x_j^{\ell}}  d v_j, \\
&\phantom{=} + \sum_{\vert \bm a \vert  = \ell + 1} \frac{b_n^{\ell + 1}}{(\ell + 1)!} \int_{[-1, 1]^q} K(\bm v) \bm v^{\bm a} D^{\vert \bm a \vert}_{\bm x} f(\bm z, \bm x+ \tau_{\bm a}\bm v) d\bm v 
\end{align*}
for some $\tau_{\bm a} \in [0, 1]$, where the second equality is due to K2. The second sum is $o(b_n^\ell)$ because all terms are bounded by A1 and K1. In summary,
\begin{align*}
	\E\bigl\{ \wt f(\bm z, \bm x) \bigr\} - f(\bm z, \bm x) &= \frac{b_n^\ell \sigma_\ell}{\ell!} \sum_{j = 1}^q \frac{\partial^\ell f(\bm z, \bm x )}{\partial x_j^\ell} + o(b_n^\ell), 
\end{align*}
as claimed.

For the variance, we get
\begin{align*}
  \var\bigl\{\wt f(\bm z, \bm x)\bigr \}  
 &= \frac{1}{nh_n^{2p} b_n^{2q}} \var\biggl\{   K\biggl(\frac{\bm Z + \bm E - \bm z}{h_{n}}\biggr)  K\biggl(\frac{\bm X -  \bm x}{b_{n}}\biggr)  \biggr\} \\
     &= \frac{1}{n} \biggl[ \frac{1}{h_n^{2p} b_n^{2q}} \E\biggl\{   K\biggl(\frac{\bm Z + \bm E - \bm z}{h_{n}}\biggr)^2  K\biggl(\frac{\bm X -  \bm x}{b_{n}}\biggr)^2 \biggr\}  \\ 
     &\phantom{= \frac{1}{n} \biggl[} - \frac{1}{h_n^{2p} b_n^{2q}} \E\biggl\{   K\biggl(\frac{\bm Z + \bm E - \bm z}{h_{n}}\biggr)  K\biggl(\frac{\bm X -  \bm x}{b_{n}}\biggr) \biggr\}^2  \biggr].
\end{align*}
The second term in square brackets has already been calculated for the bias. Using similar arguments, we can show
\begin{align*}
&\phantom{=}\; \frac{1}{n h_n^{p} b_n^{q}} \int_{\R^{p + q}}   K^2\biggl(\frac{\bm s - \bm z}{h_{n}}\biggr)  K^2\biggl(\frac{\bm t -  \bm x}{b_{n}}\biggr) f_\eta(\bm s, \bm t) d \bm s d\bm t \\
&= \kappa^p \int_{[-1,1]^q} K^2(\bm v) f(\bm z , \bm x + b_n \bm v ) d\bm v  \\
&= \kappa^{p + q} f(\bm z , \bm x)  +  o(1).
\end{align*}
Together, 
\begin{align*}
 \var\bigl\{\wt f(\bm z, \bm x)\bigr \} &= \frac{\kappa^{p + q}}{nh_n^{p} b_n^q} f(\bm z , \bm x) + \frac{f^2(\bm z , \bm x)}{n} + o\biggl( \frac{1}{nh_n^qb_n^q} \biggr) \\
& = \frac{f(\bm z, \bm x)}{n n_n^pb_{n}^q} 
     \bigl\{h_n^{-p} \kappa^{p + q} -  b_{n}^q f(\bm z, \bm x)\bigr\} + o\biggl(\frac{1}{n h_n^pb_{n}^q}\biggr).
\end{align*}

To show that the estimator is asymptotically normal, define 
\begin{align*}
Y_{i,n} = \frac{1}{nb_n^q} K\biggl(\frac{\bm Z_i + \bm E_i - \bm z}{h_n}\biggr)K\biggl(\frac{\bm X_i - \bm x}{b_n}\biggr).
\end{align*}
Then $\wt f(\bm z, \bm x) = \sum_{i = 1}^n Y_{i,n}$. which is asymptotically normal if the Lyapunov condition,
\begin{align*}
\biggl\{ \sum_{i = 1}^n \E\bigl(\vert Y_{i, n}\vert^3 \bigr) \biggr\}^{1/3}  \biggl\{ \sum_{i = 1}^n \var(Y_{i, n}) \biggr\}^{-1/2} \to 0,
\end{align*}
is fulfilled. With arguments similar to the derivation of $\var\{\wt f(\bm z, \bm x)\}$, we get $\E(\vert Y_{i, n}\vert^3) = O(n^{-1}h_n^{-2p}b_n^{-2q})$ and $\var(Y_{i, n}) = O(h_n^{-p}b_n^{-q})$. Thus,
\begin{align*}
\biggl\{ \sum_{i = 1}^n \E\bigl(\vert Y_{i, n}\vert^3 \bigr) \biggr\}^{1/3}  \biggl\{ \sum_{i = 1}^n \var(Y_{i, n}) \biggr\}^{-1/2} = O\bigl\{(nh_n^pb_n^q)^{-1/6}\bigr\},
\end{align*}
which is $o(1)$ due to assumption A4.


\subsection{Proof of Theorem 2}  \label{appendix:th2}

From the triangle inequality, we get the bound
\begin{align} \label{eq:triang}
 \bigl\vert \wt f(\bm z, \bm x) -  f(\bm z, \bm x) \bigr\vert \le  \bigl\vert \E \{\wt f(\bm z, \bm x) \} -  f(\bm z, \bm x) \bigr\vert +
 \bigl\vert \wt f(\bm z, \bm x) - \E \{\wt f(\bm z, \bm x) \}\bigr\vert.
\end{align}
We start as in the proof of \autoref{thm:distribution}, but expand \eqref{eq:kde_bias} as a Taylor polynomial of order $\ell - 2$. We can then show that for some $\tau \in [0, 1]$,
\begin{align*}
&\phantom{=} \E\bigl\{ \wt f(\bm z, \bm x) \bigr\} - f(\bm z, \bm x) \\ 
&= \frac{b_n^{\ell - 1}}{(\ell - 1) !}\sum_{j = 1}^q \int_{[-1, 1]}  K(v_j) v_j^{\ell - 1} \frac{\partial^{\ell - 1} f(\bm z, \bm x + \tau b_n \bm v )}{\partial x_j^{\ell - 1}}  d v_j \\
&= \frac{b_n^{\ell - 1}}{(\ell - 1) !}\sum_{j = 1}^q \int_{[-1, 1]}  K(v_j) v_j^{\ell - 1} \biggl\{ \frac{\partial^{\ell - 1} f(\bm z, \bm x + \tau b_n \bm v )}{\partial x_j^{\ell - 1}} -  \frac{\partial^{\ell - 1} f(\bm z, \bm x)}{\partial x_j^{\ell - 1}} \biggr\} d v_j,
\end{align*}
where the second equality holds because of K2. Using $\mathrm{A1^\prime}$, we get
\begin{align} \label{eq:uniform_bias}
\sup_{(\bm z, \bm x) \in \mathcal{S}} \bigl\vert \E\bigl\{ \wt f(\bm z, \bm x) \bigr\} - f(\bm z, \bm x) \bigr\vert 
\le  \frac{b_n^{\ell} L \tau}{(\ell -1) !}\sum_{j = 1}^q \int_{[-1, 1]} \vert K(v_j) \vert \vert v_j \vert^{\ell}d v_j  = O(b_n^\ell),
\end{align}
for a positive constant $L < \infty$. Furthermore,
\begin{align*}
 \E\bigl\{\bigl\vert \wt f(\bm z, \bm x) - \E \{\wt f(\bm z, \bm x) \}\bigr\vert^2\bigr\} = \var\bigl\{\wt f(\bm z, \bm x)\bigr \} = O\bigr\{(n h_n^{p} b_{n}^{q})^{-1}\bigl\},
\end{align*}
as in \autoref{thm:distribution}. And since convergence in $L^2$ implies convergence in probability, 
\begin{align*}
\bigl\vert \wt f(\bm z, \bm x) - \E \{\wt f(\bm z, \bm x) \}\bigr\vert = O_p\bigr\{(n h_n^{p} b_{n}^{q})^{-1/2}\bigl\},
\end{align*}
which, together with \eqref{eq:uniform_bias}, proves \eqref{eq:weak_consistency}.

Moreover, there is a positive constant $\overline c_1 < \infty$ such that almost surely
\begin{align} \label{eq:einmahl_mason}
\lim_{n \to \infty} \sqrt{\frac{n h_n^pb_n^q}{\max\{\ln \ln n, \ln h_n^{-1}, \ln b_n^{-1}\}}}  \sup_{\mathcal{S}} \bigl\vert \wt f(\bm z, \bm x) -  \E \{\wt f(\bm z, \bm x) \} \bigr\vert \le \overline c_1,
\end{align}
see Theorem 1 of \citet{Einmahl05}. Combining \eqref{eq:uniform_bias} and \eqref{eq:einmahl_mason} proves \eqref{eq:strong_consistency}.


\subsection{Proof of Theorem 3}  \label{appendix:th3}

Note that we can write
\begin{align*}
\E\bigl\{d^2(\wt f, f)\bigr\} = \E\bigl\{\sup_{\mathcal{S}^\prime} \bigl\vert \wt f(\bm z, \bm x) - f(\bm z, \bm x) \bigr\vert^2 \bigr\},
\end{align*}
where $\mathcal{S}^\prime = \{(\bm z, \bm x)\}$ for $d_{(\bm z, \bm x)}$ and $\mathcal{S}^\prime = \mathcal{S}$ for $d_{\infty, \mathcal{S}}$. 
It holds
\begin{align}
\frac 1 2 \E\bigl\{d^2(\wt f, f)\bigr\} &\le \sup_{\mathcal{S}^\prime} \bigl\vert \E \{\wt f(\bm z, \bm x) \} -  f(\bm z, \bm x) \bigr\vert^2 +
\E\bigl[ \sup_{\mathcal{S}^\prime} \bigl\vert \wt f(\bm z, \bm x) - \E \{\wt f(\bm z, \bm x) \}\bigr\vert^2 \bigr] \notag \\
&= a_1 + a_2 \label{eq:bias_variance}
\end{align}
Using arguments almost identical to \eqref{eq:uniform_bias}, we obtain
\begin{align*}
\sup_{(\bm z, \bm x) \in \mathcal{S}} \bigl\vert \E\bigl\{ \wt f(\bm z, \bm x) \bigr\} - f(\bm z, \bm x) \bigr\vert
\le  \frac{b_n^{\beta} \lambda \tau^{\beta - r} }{r !}\sum_{j = 1}^q \int_{[-1, 1]} \vert K(v_j) \vert \vert v_j \vert^{\beta}d v_j  = b_n^\beta \overline c_2,
\end{align*}

For bounding $a_2$, we need to consider the characteristics of scenarios (i)--(iv). 
\begin{enumerate}
\item We proceed as in the proof of \autoref{thm:distribution} to get 
\begin{align*}
a_2 = \var\bigl\{\wt f(\bm z, \bm x)\bigr \} &= \frac{\kappa^{p + q}}{nh_n^{p} b_n^q} f(\bm z , \bm x) + \frac{f^2(\bm z , \bm x)}{n} + o\biggl( \frac{1}{nb_n^q} \biggr).
\end{align*}
For $q \ge 1$, choosing $b_n \sim n^{-1 / (2\beta + q)}$ yields
\begin{align*}
\limsup_{n \to \infty} n^{2\beta/(2\beta + q)} a_2 \le \frac{\kappa^{p + q}}{h_0^{p}}f(\bm z , \bm x) = \overline c_3 < \infty.
\end{align*}
If $q = 0$, it holds $f \le 1$, and we get 
\begin{align}
\limsup_{n \to \infty}  n a_2 \le \frac{\kappa^{p + q}}{h_0^{p}}f(\bm z) + f^2(\bm z)  \le \frac{\kappa^{p + q}}{h_0^{p}} + 1 = \overline c_4 < \infty. \label{eq:i}
\end{align}

\item With $h_n \sim 1$ and $b_n \sim (n/\ln n)^{-1/(2\beta + q)}$ in \eqref{eq:einmahl_mason}, we get
\begin{align*}
 \limsup_{n \to \infty}  (n/\ln n)^{2\beta/(2\beta + q)} a_2 \le \overline c_1,
\end{align*}

\item Using \eqref{eq:i} yields
\begin{align*}
\limsup_{n \to \infty}  n a_2 \le \sum_{\bm z \in \mathcal{S}^\prime} \limsup_{n \to \infty}  \E\bigl[\bigl\vert \wt f_{\bm Z}(\bm z) - \E \{\wt f_{\bm Z}(\bm z) \}\bigr\vert^2 \bigr]  
\le \vert \mathcal{S}^\prime \vert \overline c_4 = \overline c_5 < \infty.
\end{align*}

\item With $h_n \sim 1$ and $b_n = 1$ in \eqref{eq:einmahl_mason}, we get
\begin{align*}
\limsup_{n \to \infty}  (n/\ln \ln n) a_2 \le \overline c_1.
\end{align*}
\end{enumerate}

Setting $\overline c = 2(\overline c_1 + \overline c_2 + \overline c_3 + \overline c_4 + \overline c_5)$ concludes the proof.


\subsection{Proof of Theorem 4} \label{appendix:th4}

We start with lower bounds for (i) and (iii). Fix $\bm z \in \Z^p$ and define 
\begin{align*}
\mathcal{G}_{1}(\beta, \lambda) =\bigl\{f \in \mathcal{H}(\beta, \lambda) \colon f(\bm z^\prime, \bm x) = 0 \mbox{ for } \bm z^\prime \neq \bm z \bigr\}.
\end{align*}
This set contains all probability densities in $\mathcal{H}(\beta, \lambda)$ that correspond to a random vector $(\bm Z, \bm X)$ with $\bm Z = \bm z$ almost surely. This is equivalent to the case where all variables are continuous. By definition, $\mathcal{G}_{1}(\beta, \lambda) \subset  \mathcal{H}(\beta, \lambda)$ and, thus, $\mathcal{R}_n^*\{\mathcal{G}_{1}(\beta, \lambda), d\} \le \mathcal{R}_n^*\{\mathcal{H}(\beta, \lambda), d\}$. The  two rates in \autoref{thm:lower} (i) and (iii) then follow from Theorem 9 in \citep{ibragimov1983}. 

For (ii) and (iv),  we can simply consider a parametric family of densities $\mathcal{G}_2$. This yields the classical lower bound $n^{-1/2}$ for estimating a  finite dimensional parameter \citep[see, e.g.,][Chapter 2]{tsybakov2008}.

	\bibliography{references}
		
\end{document}